\documentclass[a4paper,fleqn,usenatbib]{mnras}

\usepackage[T1]{fontenc}
\usepackage{ae,aecompl}

\usepackage{graphicx}	
\usepackage{amsmath}	
\usepackage{amssymb}	
\usepackage{relsize}

\def\INSPIRE{\mbox{{\tt INSPIRE}}}
\newcommand{\Reff}{$\mathrm{R}_{\mathrm{e}\,}$}
\newcommand{\Mstar}{M$_{\star}\,$}

\newcommand{\kms}{km s$^{-1}$}
\newcommand{\Msun}{M$_{\odot}\,$}

\usepackage[normalem]{ulem}
\newcommand{\ppxf}{\textsc{pPXF}}

\defcitealias{Tortora+18_UCMGs}{T18}
\defcitealias{Scognamiglio20}{S20}
\defcitealias{Ferre-Mateu+17}{F17}
\defcitealias{Spiniello20_Pilot}{\INSPIRE\, Pilot}
\defcitealias{Spiniello+21}{\INSPIRE\, DR1}

\title[The IMF in relic galaxies]{INSPIRE: INvestigating Stellar Population In RElics IV. \\ The Initial Mass Function slope in relics}

\author[I. Mart\'in-Navarro]{Ignacio Mart\'in-Navarro$^{1,2}$\thanks{E-mail: imartin@iac.es}, C. Spiniello$^{3,4}$, C. Tortora$^{4}$, L. Coccato$^{5}$, G. D'Ago$^{6}$, \and A. Ferr\'e-Mateu$^{1,2}$,  C. Pulsoni$^{7}$, J. Hartke$^{5,3,8}$, M.~ Arnaboldi$^{5}$, L. Hunt$^{9}$, \and N.~R.~Napolitano$^{10}$, D. Scognamiglio$^{11}$ and M. Spavone$^{4}$  \\ 
$^{1}$Instituto de Astrof\'isica de Canarias, V\'ia L\'actea s/n, E-38205 La Laguna, Tenerife, Spain\\
$^{2}$Departamento de Astrof\'isica, Universidad de La Laguna, E-38205 La Laguna, Tenerife, Spain\\
$^{3}$Sub-Dep. of Astrophysics, Dep. of Physics, University of Oxford, Denys Wilkinson Building, Keble Road, Oxford OX1 3RH, UK\\
$^{4}$INAF -  Osservatorio Astronomico di Capodimonte, Via Moiariello  16, 80131, Naples, Italy\\
$^{5}$European Southern Observatory, Alonso de Córdova 3107, Santiago de Chile, Chile\\
$^{6}$Pontificia Universidad Cat\'olica de Chile, Av. Vicu\~na Mackenna 4860, 7820436 Macul, Santiago, Chile\\
$^{7}$Max-Planck-Institut f\"{u}r  extraterrestrische Physik, Giessenbachstrasse, 85748 Garching, Germany\\
$^{8}$Finnish Centre for Astronomy with ESO (FINCA), University of Turku, FI-20014 Turku, Finland \\
$^{9}$ INAF - Osservatorio Astronomico di Arcetri, Largo Enrico Fermi 5, 50125, Firenze, Italy\\
$^{10}$School for Physics and Astronomy, Sun Yat-sen University, Guangzhou 519082, Zhuhai Campus, China\\
$^{11}$Argelander-Institut f\"{u}r Astronomie, Auf dem H\"{u}gel 71, D-53121 - Bonn, Germany \\
}

\date{Accepted XXX. Received YYY; in original form ZZZ}

\pubyear{2022}

\begin{document}
\label{firstpage}
\pagerange{\pageref{firstpage}--\pageref{lastpage}}
\maketitle

\begin{abstract}
In the last decade, growing evidence has emerged supporting a non-universal stellar Initial Mass Function (IMF) in massive galaxies, with a larger number of dwarf stars with respect to the Milky-Way  (bottom-heavy IMF). However, a consensus about the mechanisms that cause IMF variations is yet to be reached. Recently, it has been suggested that stars formed early-on in cosmic time, via a star formation burst, could be characterised by a bottom-heavy IMF.  A promising  way to confirm this is to use relics, ultra-compact massive galaxies, almost entirely composed by these ”pristine” stars. The INSPIRE Project aims at assembling a large sample of confirmed relics, that can serve as laboratory to investigate on the conditions of star formation in the first 1-3 Gyr of the Universe. In this third INSPIRE paper, we build a high signal-to-noise spectrum from five relics and one from five galaxies with similar sizes, masses, and kinematical properties, but characterised by a more extended star formation history (non-relics). Our detailed stellar population analysis suggests a systematically bottom-heavier IMF slope for relics than for non-relics, adding new observational evidence for the non-universality of the IMF at various redshifts and further supporting the above proposed physical scenario.

\end{abstract}




\section{Introduction}
Relic galaxies are defined as ultra-compact massive galaxies that formed almost the totality of their stellar masses at very high redshift through a short and intense star formation burst (star formation declining time $\sim100$ Myr, star formation rate $\ge1000$ M$_{\odot}\, yr^{-1} $), and then evolved undisturbed without experiencing any mergers or interactions until the present day \citep{Trujillo+14}\footnote{Within the INSPIRE project, we use the definition of \citet{Trujillo+09_superdense} which sets the following thresholds in stellar mass and size to define relics: \Mstar$>6\times10^{10}$\Msun and \Reff$<2$ kpc}. Hence, they provide a unique opportunity to study the mechanisms of star formation at high-z. However, relics are extremely rare and, to-date, only a handful have been spectroscopically confirmed \citep{Ferre-Mateu+17}.  

The INvestigating Stellar Population In Relics (\INSPIRE) Project \citep{Spiniello20_Pilot, Spiniello+21} aims at building the first large catalogue of relic galaxies at $0.1<z<0.5$ to put constraints on the first phase of the mass assembly of massive Early-Type Galaxies (ETGs) in the Universe, and on the mechanisms responsible for their size evolution over cosmic time \citep[e.g.,][]{Buitrago+18_compacts}.  
In \citet[][hereafter \INSPIRE\, DR1]{Spiniello+21}, thanks to high UVB+VIS signal-to-noise (SNR) X-Shooter@VLT (XSH) spectra, we have constrained in detail the stellar population parameters of 19 spectroscopically confirmed ultra-compact massive galaxies (UCMGs, from \citealt{Tortora+18_UCMGs} and \citealt{Scognamiglio20}). We first inferred light-weighted [Mg/Fe] from line-index strengths, and then computed mass-weighted mean age and [M/H], performing full-spectral fitting on the spectra with the Penalised Pixel-fitting software (\ppxf; \citealt{Cappellari04,Cappellari17}). We found that 10 objects have formed more than $75\%$ of their stellar mass (M$_{\star}$) within 3 Gyr from the Big Bang ($M_{\star, t_{3 \text{Gyrs}}}$), and hence classify them as relics, increasing by a factor of 3.3 the current number of spectroscopically confirmed relics\footnote{In DR1, we identify 4 `extreme relics' with $M_{\star, t_{3 \text{Gyrs}}}\sim100$\%. Here, for simplicity, we only separate the systems in two families.}. 
The remaining 9 UCMGs showed instead a more extended star formation history (SFH), despite being overall old ($>5$ Gyrs). 

In this paper, we investigate whether the stellar Initial Mass Function (IMF) slope differs between relics and non relics. This idea is motivated by a growing number of observations finding a dwarf-rich IMF slope in the centres of massive galaxies, where the pristine, oldest bulk of the stellar population is expected to dominate  \citep[e.g.][]{Martin-Navarro+15_IMF_variation,Sarzi+18,LaBarbera+19, Barbosa21}. Theoretical works 
predict that the extreme star formation conditions (i.e., higher temperature and density, hence higher Mach numbers, \citealt{Hennebelle08}) under which these stars (that constitute the great majority of stars in relics) formed could have favoured the fragmentation of molecular clouds, resulting in a non-universal IMF with an excess of low-mass stars \citep[e.g. ][]{Chabrier_2014}. This hypothesis has been supported by the direct inference of the IMF slope in the only three local confirmed relics where the IMF has been measured \citep{Martin-Navarro+15_IMF_relic,Ferre-Mateu+17}.  Thanks to \INSPIRE\, we can now extend the investigation to higher redshifts and to relatively larger number statistics.

\begin{figure}
\includegraphics[width=8.5cm]{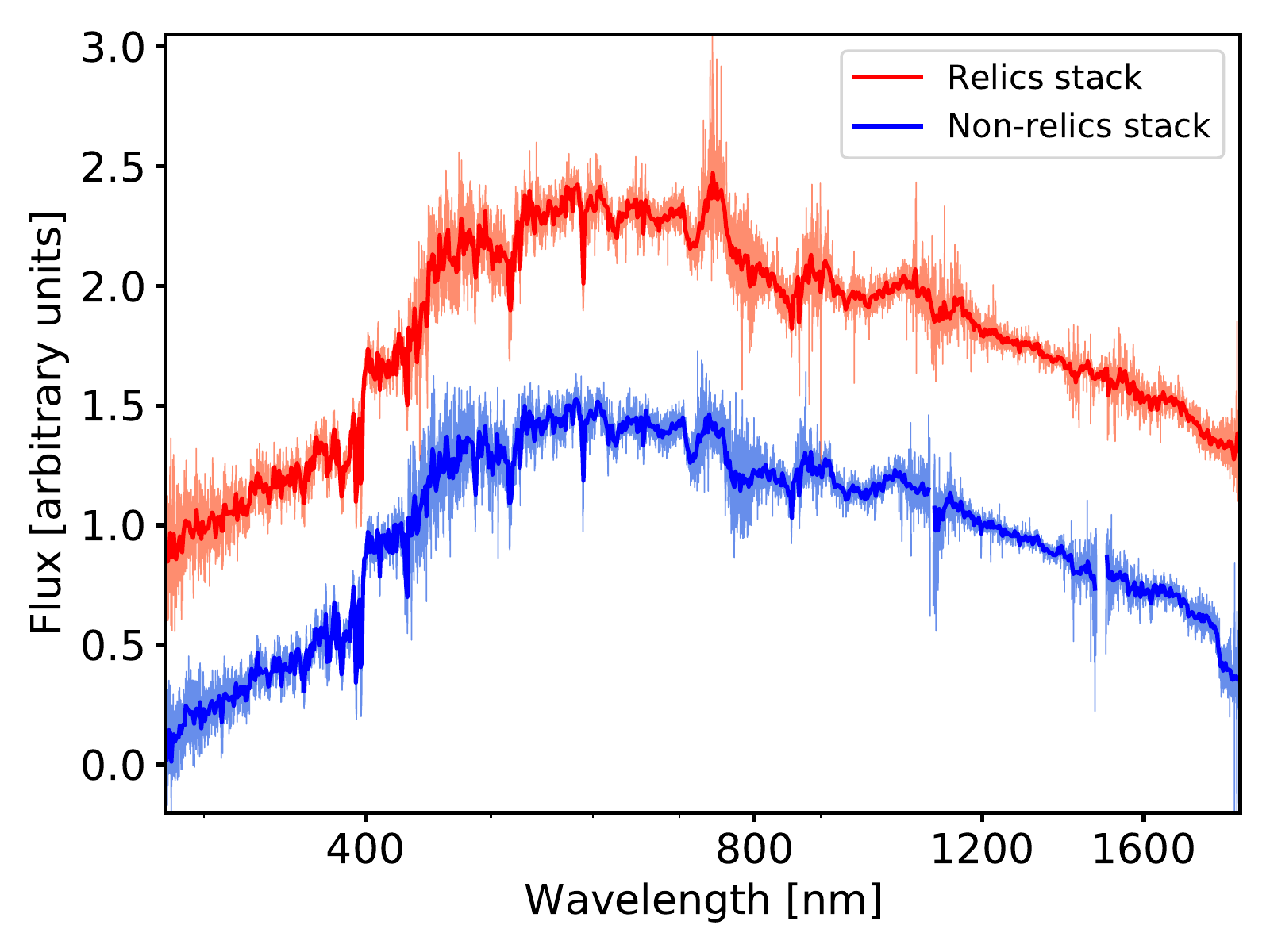}
\caption{Stacked 1D spectra for relics (red) and non relics (blue). The spectra at the original resolution are plotted as light lines, while darker lines show a smoothed version to highlight the most prominent stellar absorption features.}
\label{fig:spectra}
\end{figure}

\begin{figure}
\includegraphics[width=8.5cm]{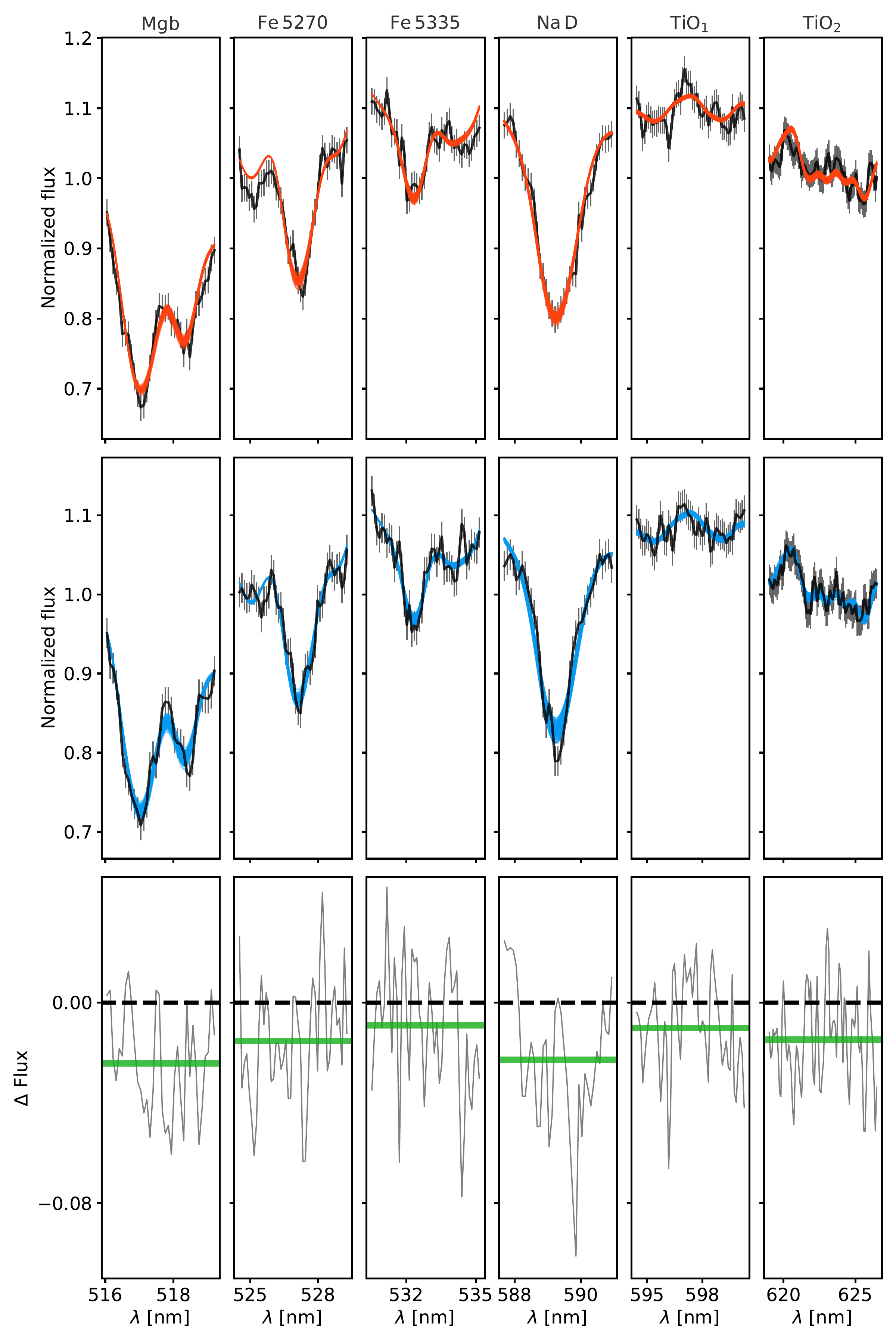}
\caption{XSH data for relics (top) and non-relics (middle) spectra compared to the MILES best-fitting models (50 samples drawn from the posterior distribution, shown in colored lines). Each horizontal sub-panel shows the band-pass of one of the spectral indices included in our analysis. MILES models have been convolved to match the resolution of the data. The bottom panel shows the continuum-corrected differences between the spectra of relic and non-relic galaxies. Green horizontal lines indicate the average difference between the two spectra, which is negative for all features.}
\label{fig:fitting}
\end{figure}

\section{The data: stacks by `relic families'}
Constraining the IMF slope directly from spectral fitting is a challenging task. It requires very high SNR spectra, covering a wavelength range that is large enough to break the degeneracy between IMF variation and variation of  other stellar population parameters \citep{Spiniello+14, Spiniello+2015}. 

Unfortunately, the INSPIRE DR1 spectra of individual galaxies do not reach such SNRs (see Table C1 in \citetalias{Spiniello+21}) and/or are affected by sky residuals, bad pixels, and other systematics. Therefore we need to stack the spectra to reach the necessary SNR and spectral cleanliness to enable IMF studies. 
Since we aim at study the effect of the `relicness'/star formation conditions on the observed IMF, we need to fix the other physical parameters which have been proposed as possible drivers of IMF variations (e.g., stellar velocity dispersion, metallicity, and/or [$\alpha$/Fe],  \citealt{TRN13_SPIDER_IMF,Spiniello+14,McDermid+14_IMF,Martin-Navarro15_CALIFA}), but that do not fully explain some observations \citep[e.g.,][]{Barbosa21_letter,Martin-Mavarro+21}. 
Our strategy is to stack together 5 relics (J0317-2957, J0838+0052, J0847+0112, J0920+0212, J2359-3320) and 5 non-relics (J0226-3158, J0240-3141, J0314-3215, J0321-3213, J0326-3303) from \citetalias{Spiniello+21}. This is the maximum number of systems per `relic family' that we can stack in order to have the two final 1D stacked spectra with almost identical stellar velocity dispersion ($\sigma$), and very similar metallicity ([M/H]) and [Mg/Fe], to ensure that the resulting trends are not simply driven by differences in velocity dispersion, stellar mass, or other stellar population parameters. The sizes and stellar masses of the individual systems are listed in Table~\ref{tab:individual_prop}. Effective radii are the median values obtained from the values computed in $g,r,i$ optical bands. They have been then translated into kpc using the Python version of the Ned Wright's Cosmology Calculator \citep{Wright06} \footnote{\url{http://www.astro.ucla.edu/~wright/CosmoCalc.html}}. Stellar masses have been computed via SED-fitting of $ugri$ bands \citep{Tortora+18_UCMGs,Scognamiglio20}. In the same table, we also provide (last column) the fraction of stellar mass that was assembled by $z=2$, corresponding to 3 Gyr after the Big Bang \citep[assumed here to be the end of the contraction phase, see e.g.][]{Zolotov15}. Relics (upper block) were almost completely assembled by then, while non-relics (lower block) were already at an advanced stage of their stellar mass assembly (M$_{\star, z=2}$ >60\% for all systems) but did not complete it yet. Thus, this delayed mass assembly appears as the only significant difference separating relics and non-relics.

\begin{table} 
\begin{tabular}{|l|cccc|}
\hline
  \multicolumn{1}{|c|}{ID} &
  \multicolumn{1}{c}{$<R_{\text{eff}}>$} &
  \multicolumn{1}{c}{$<R_{\text{eff}}>$} &
  \multicolumn{1}{c}{M$_{\star}$} &
\multicolumn{1}{c}{M$_{\star,z=2}$} \\

\multicolumn{1}{|c|}{KiDS} &
  \multicolumn{1}{c}{$(\arcsec)$} &
  \multicolumn{1}{c}{(kpc)} &
  \multicolumn{1}{c}{$(10^{11}M_{\odot})$}   &
    \multicolumn{1}{c|}{(\%)} \\
\hline
 &   &  Relics  &   &  \\
\hline
J0317-2957 &  0.26  & 1.05  & 0.87 & 94.5\\
J0838+0052 &  0.31  & 1.28  & 0.87 & 95.5\\
J0847+0112 &  0.46  & 1.37  & 0.99 &  100\\
J0920+0212 &  0.34  & 1.48  & 1.03 &  96.0\\
J2359-3320 &  0.24  & 1.04  & 1.07 &  97.5\\
\hline
 &   &  Non-Relics  &   &  \\
\hline
J0226-3158 &  0.35  & 1.32  & 0.69 &  68.5\\
J0240-3141 & 0.19  & 0.81  & 0.98 &  64.5\\
J0314-3215  & 0.15  & 0.66  & 1.00 &  62.0\\
J0321-3213  & 0.31  & 1.37  & 1.23 &  64.5\\
J0326-3303  & 0.32  & 1.44  & 0.93 &  73.5\\
\hline
\end{tabular}
\caption{Sizes and stellar masses of the individual \INSPIRE\ systems, computed in \citetalias{Tortora+18_UCMGs} and \citetalias{Scognamiglio20} from KiDS photometry. The last column lists the percentage of stellar mass already in place 3 Gyr after the Big Bang (\citetalias{Spiniello+21}). }
\label{tab:individual_prop}
\end{table}

The kinematic and (mass-weighted) stellar populations properties of the selected objects can instead be found in \citetalias{Spiniello+21} (Table~3 and Table~4).  Figure~\ref{fig:spectra} shows the final  SNR ($\sim 60$ per \AA) stacked spectra on which we run our stellar population fitting code, as described below. 

\begin{table*}
\begin{tabular}{|l|ccc|cccc|}
\hline

  \multicolumn{1}{|l|}{STACKED} &
  \multicolumn{1}{|c|}{$\sigma$ [km\,s$^{-1}$]} &
  \multicolumn{1}{c|}{Age [Gyr]} &
  \multicolumn{1}{|c|}{$\Gamma_\mathrm{B}$} &  
  \multicolumn{1}{c|}{[M/H] } &
  \multicolumn{1}{c|}{[Mg/Fe] } &
  \multicolumn{1}{c|}{[Ti/Fe] } &
  \multicolumn{1}{c|}{[Na/Fe]} \\
\hline
Relics & $213\pm 5$ & $12.2\pm 0.9$ & $2.4\pm 0.2$ & $+0.05\pm 0.02$ &  $0.17\pm 0.03$ &  $0.20\pm 0.10$ & $0.10\pm 0.05$ \\ 
Non-Relics &  $210\pm 4$ & $7.1\pm 1.3$ & $1.2\pm 0.6$ & $-0.01\pm 0.04$ & $0.14\pm 0.03$ & $0.05\pm 0.15$ & $0.10\pm 0.05$ \\ 
\hline
\end{tabular}                      
\caption{
Results of the stellar population analysis on the stacked spectra for relics and non relics, according to our fiducial model. From left to right: stellar velocity dispersion and age computed via full spectra fitting, and metallicity, IMF slope, Mg, Ti and Na abundances, computed via full index fitting. In addition to the formal errors on the velocity dispersion, a systematic uncertainty of $\sim$20 \kms \ is expected depending on the details of the fitting scheme (INSPIRE DR2, D’Ago et al. 2022, submitted).}
\label{tab:stack_properties}
\end{table*}

\section{Stellar population analysis}
For the stellar population analysis we use the Full Index Fitting (FIF) technique, tested and described in \citet{Martin-Navarro+19,Martin-Mavarro+21}. The FIF consists of two basic steps. First, we measure the kinematics of the stacked spectra and derive their luminosity-weighted ages from the best-fit linear combination of SSPs retrieved by \ppxf. We impose the regularization scheme described in \citet{Cappellari17} to overcome the ill-constrained nature of the inversion problem \citep{Ocvirk06}. We repeat the fitting process 10 times to assess the uncertainties on the estimated parameters. Second, we measure metallicity, IMF slope, and  elemental abundance ratios by fitting every pixel within the band-pass of the standard line-index definition \citep[see][]{Martin-Navarro+19}, assuming the luminosity-weighted age computed in the first step and take into account its uncertainty. Our reference set of indices comprises the following spectral features: Mgb\,5177, Fe\,5270, Fe\,5335, NaD, TiO$_1$, and TiO$_2$. Our fiducial stellar population model returns the [M/H], [Mg/Fe], [Na/Fe], [Ti/Fe] abundance ratios \footnote{Note that neither [Ti/Fe] nor [Na/Fe] abundances are self-consistently treated in the MILES models.}  and the IMF slope for each stacked spectrum. In addition, our likelihood calculation includes a correction term to assumed error spectra $\Delta_\mathrm{err}$ \citep{emcee}.  To test the robustness of the results, we repeat our analysis varying the indices and assumptions of the fitting process. The results of all these tests are described in \S~\ref{sec:results}.

The stellar population analysis is powered by the MILES stellar population synthesis models \citep{Vazdekis10,Vazdekis15}. These models cover a range in ages from 0.03 to 14 Gyr, and from $-2.27$ to $+0.26$ dex in total metallicity. We use the semi-empirical $\alpha$-variable SSP fed with the BaSTI set of isochrones \citep{basti1,basti2}, calculated at [$\alpha$/Fe]=0.4 and [$\alpha$/Fe] = 0.0, and apply the response functions of \citet{Conroy_vanDokkum12a} to model the variations of non-$\alpha$ elements. IMF variations in the MILES models are parametrized through the slope of the high-mass end $\Gamma_\mathrm{B}$ which, by normalization, effectively changes the relative fraction of low-mass stars. For reference, a Milky Way-like IMF is characterized by a slope $\Gamma_\mathrm{B} \sim 1.3$.

Figure~\ref{fig:fitting} presents the result of our fitting procedure, for relics (top) and non-relics (middle). For each index, we show 50 different samples drawn from the full posterior distribution (coloured lines) overplotted on the XSH data (black). The SSP models are convolved to match the resolution of the data. The quality of both data and models allows a robust inference on the IMF slope as described in the following section. Finally, in the bottom panel we show the (continuum-corrected) difference between the relic and non-relic spectra. Green horizontal lines mark the mean difference for each spectral feature. All of the selected absorption features are always stronger in the stack of relic galaxies, which indicates distinct stellar population properties and hints already to a bottom-heavier IMF slope.

\begin{figure*}
  \includegraphics[width=8.5cm]{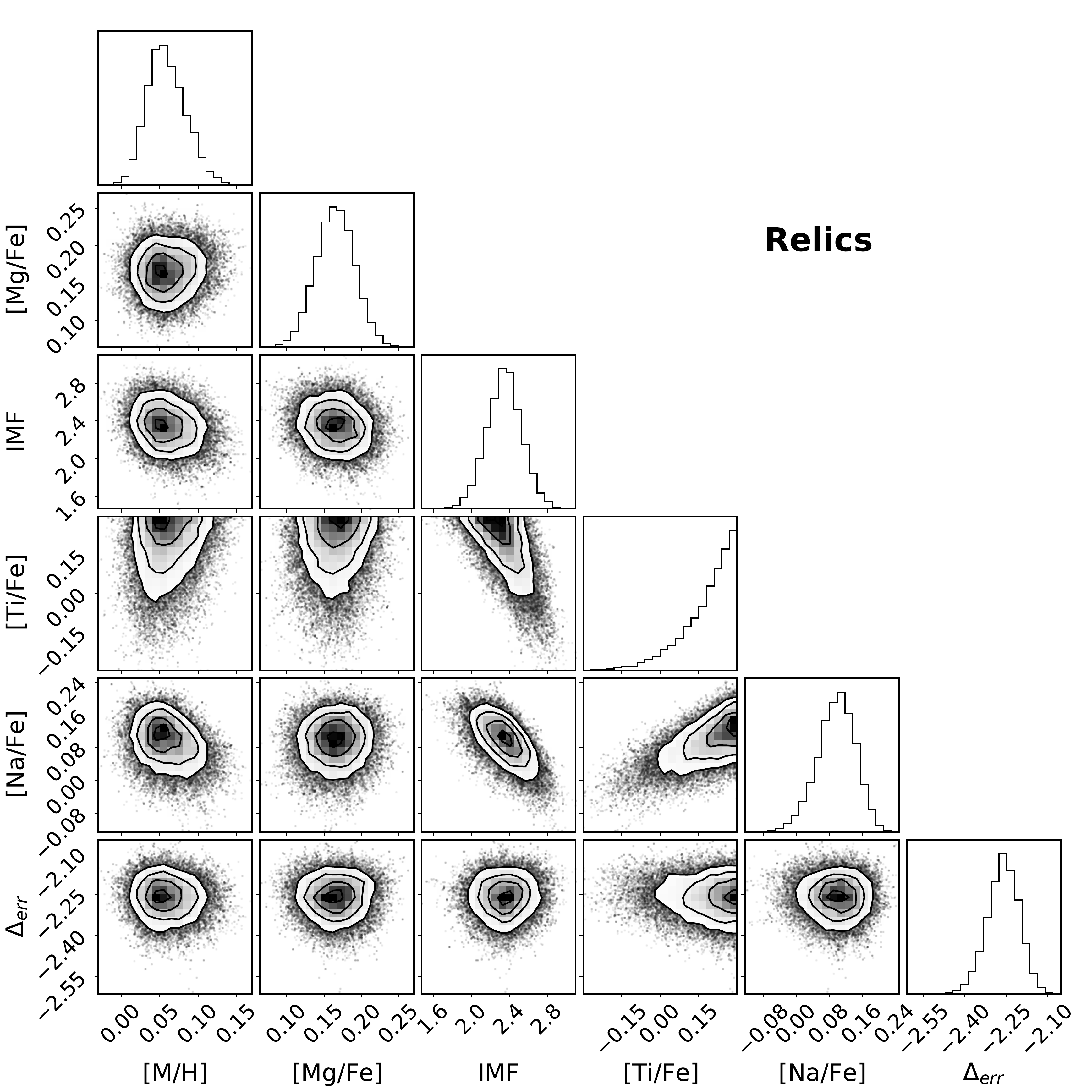}
  \includegraphics[width=8.5cm]{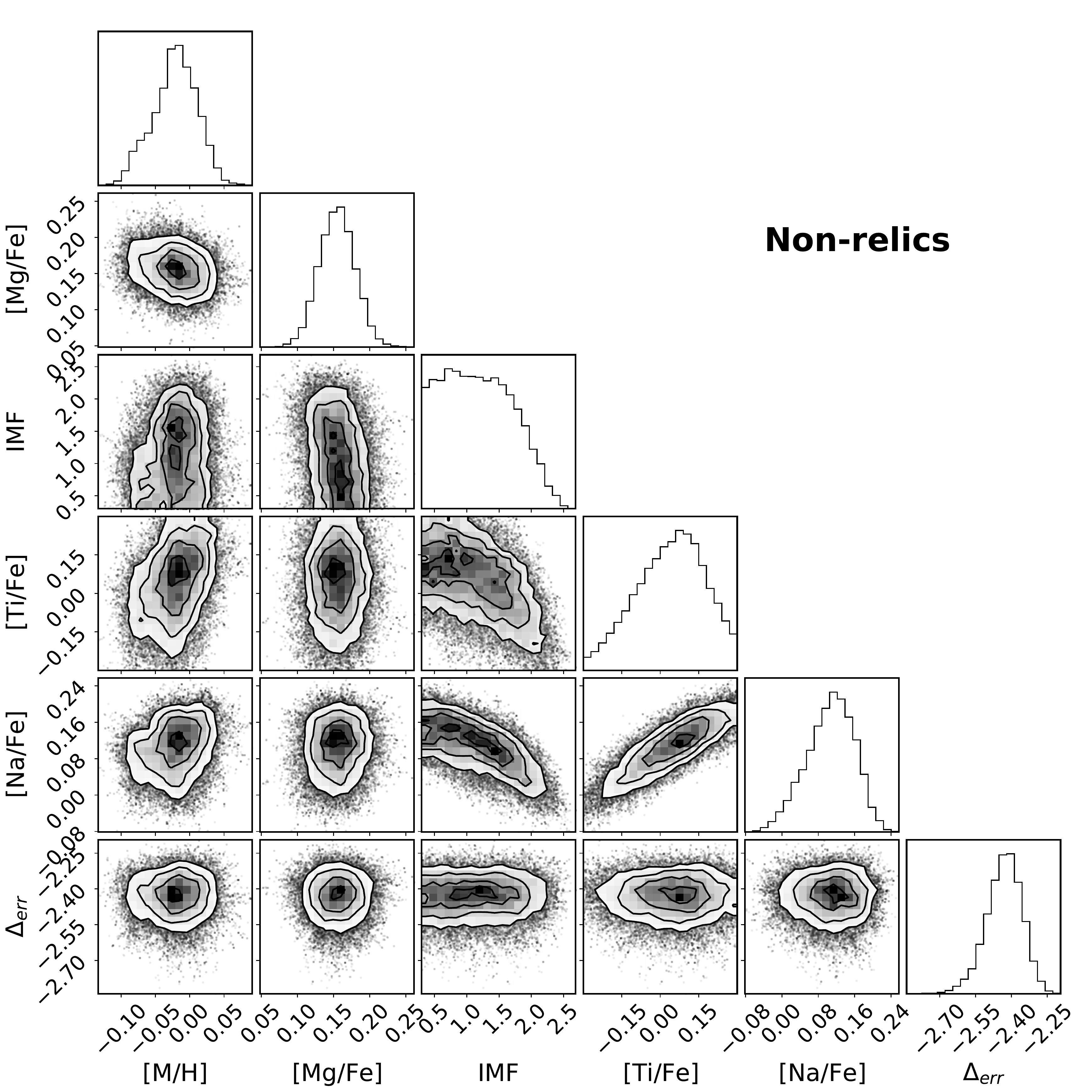}
  \caption{Full posterior distribution for the stack spectrum of relic (left) and non-relic galaxies (right) resulting from our FIF analysis. Note that the age has been fixed to the value measured using pPXF. The derived IMF slope value for the relics stack is higher (i.e. bottom-heavier) than that measured for the non-relic sample.}
  \label{fig:corner}
\end{figure*}

\section{Results} 
\label{sec:results}
The results of our stellar population analysis are summarized in Table~\ref{tab:stack_properties} and the full posterior distributions are shown in Fig.~\ref{fig:corner}. Note that the the age and metallicity, as well as the other stellar population parameters, computed here are light-weighted. In contrast age and metallicity presented in \citetalias{Spiniello+21} were mass-weighted (slightly younger ages and higher [M/H]s.) As expected from their definition, and in agreement with previous measurements, relic galaxies are significantly older than the non-relics. 

\subsection{IMF slope: relics vs non-relics}

\begin{figure}
\includegraphics[width=8.5cm]{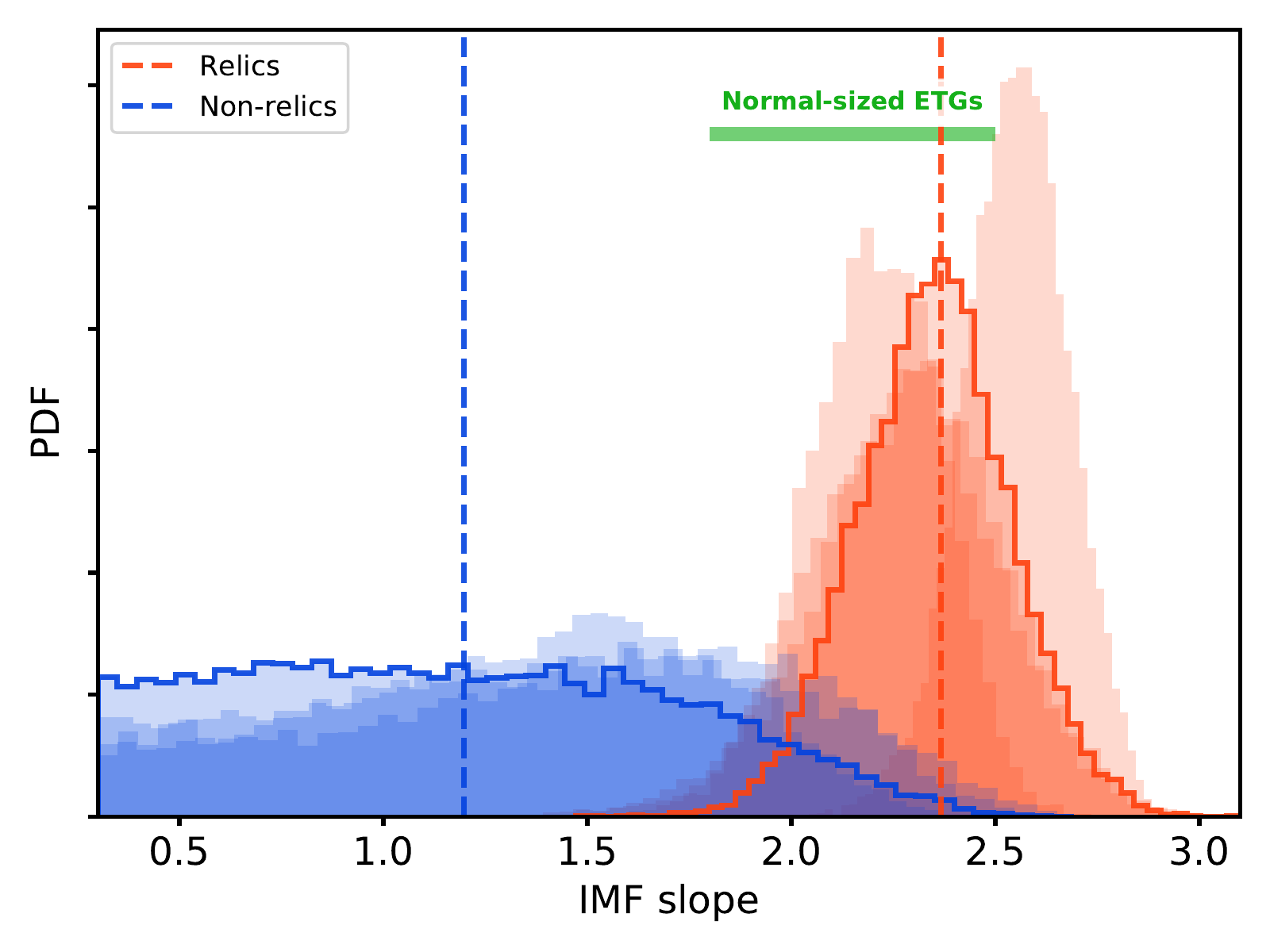}
\caption{Systematic difference in the IMF slope. Different histograms correspond to different tests varying some of our modelling assumptions (see text for more details). The PDF resulting from our fiducial model is highlighted with a solid line while vertical dashed lines indicate the median of the distribution. The IMF slopes measured for the stack of relics (red) appear systematically steeper (i.e. dwarf-dominated) than the IMF slopes of the non-relics stack (blue). The green horizontal line shows the expected IMF slope range for normal-sized ETGs for the velocity dispersion measured in both stacked spectra ($\sigma=210$ km\,s$^{-1}$, see e.g., \citealt{LaBarbera+13_SPIDERVIII_IMF,Spiniello+14}).}
\label{fig:imf_pdf}
\end{figure}

Arguably, the most interesting difference between relics and  non-relics is related to their IMF slopes. For the stack of relic galaxies the IMF is, on average, systematically more biased towards low-mass stars, in agreement with what is measured in local relics \citep{Martin-Navarro+15_IMF_relic,Ferre-Mateu+17}. 
This is best exemplified by the marginalised posterior distribution shown in Figure~\ref{fig:imf_pdf}. According to our fiducial modelling described above, the average IMF slope of relics is  $\Gamma_\mathrm{B} = 2.4\pm 0.2$, also qualitatively consistent with that predicted by the IMF-$\sigma$ relation for normal-sized galaxies (e.g., \citealt{LaBarbera+13_SPIDERVIII_IMF, Spiniello+14}).

From the stacked spectrum of the five non relics, despite having very similar velocity dispersion ($\sim210$ \kms), metallicity and [Mg/Fe] than the relic one, we infer an IMF slope consistent with that of the Milky Way for our reference stellar population modelling, although with a larger uncertainty than for the relics stack. We speculate that this might have two origins. On one side, from a technical point of view, the underlying SSP assumption of our stellar population modelling becomes less robust for systems with more extended SFHs, as it is the case for non-relics. In addition, the sensitivity to IMF variations of our set of indices is weaker in the Milky Way-like regime \citep[see e.g.][]{LaBarbera+13_SPIDERVIII_IMF}, leading to less constrained solutions. On the other side, a physical reason is to be found in the fact that, despite these 5 galaxies do not pass the operative threshold set in the \citetalias{Spiniello+21} ($M_{\star, t_{3 \text{Gyrs}}} \ge 75$\%), some of them still formed a large fraction of their stellar masses at high redshift. So, for example, J0326-3303, which formed almost 70\% of its $M_{\star}$ during the first phase of the mass assembly in the Universe, might have an integrated IMF slope steeper than the IMF of J0314-3215, which formed only $\sim60\%$ of its $M_{\star}$ at early epochs. The combination of these two factors might also explain why our stack of non-relic objects shows an IMF slope which is slightly bottom-lighter than expected from their mass.

The different shaded histograms in the figure show the probability density distributions (PDFs) resulting from the different tests we performed to assess the robustness of our measurements. In particular, we explore variations of our fiducial model by:  i) excluding [Ti/Fe] and [Na/Fe] as free parameters in the fitting process, ii) removing the NaD feature which might be affected by model systematics and absorption by neutral sodium in the interstellar medium \citep[e.g.][]{Spiniello+14}, iii) changing the level of regularization when measuring the luminosity-weighted ages. Moreover, since the [Ti/Fe] posterior distribution for the relic stack is clustered around the model boundary in Fig.~\ref{fig:corner} and this quantity is anti-correlated with the slope of the IMF, iv) we also fit the stack of relic galaxies assuming the maximum [Ti/Fe] = 0.3 allowed by our model. This last test leads to a slightly less extreme IMF, but still significantly steeper than that measured for the control stack. Allowing a wider range in [Ti/Fe] would further reduce the significance of the IMF difference between the relics and the non-relics, although given the observed abundance pattern of nearby relic galaxies \citep{Martin-Navarro+15_IMF_relic} it is unlikely that they trully exhibit much higher [Ti/Fe] ratios.

\begin{figure}
  \includegraphics[width=8.5cm]{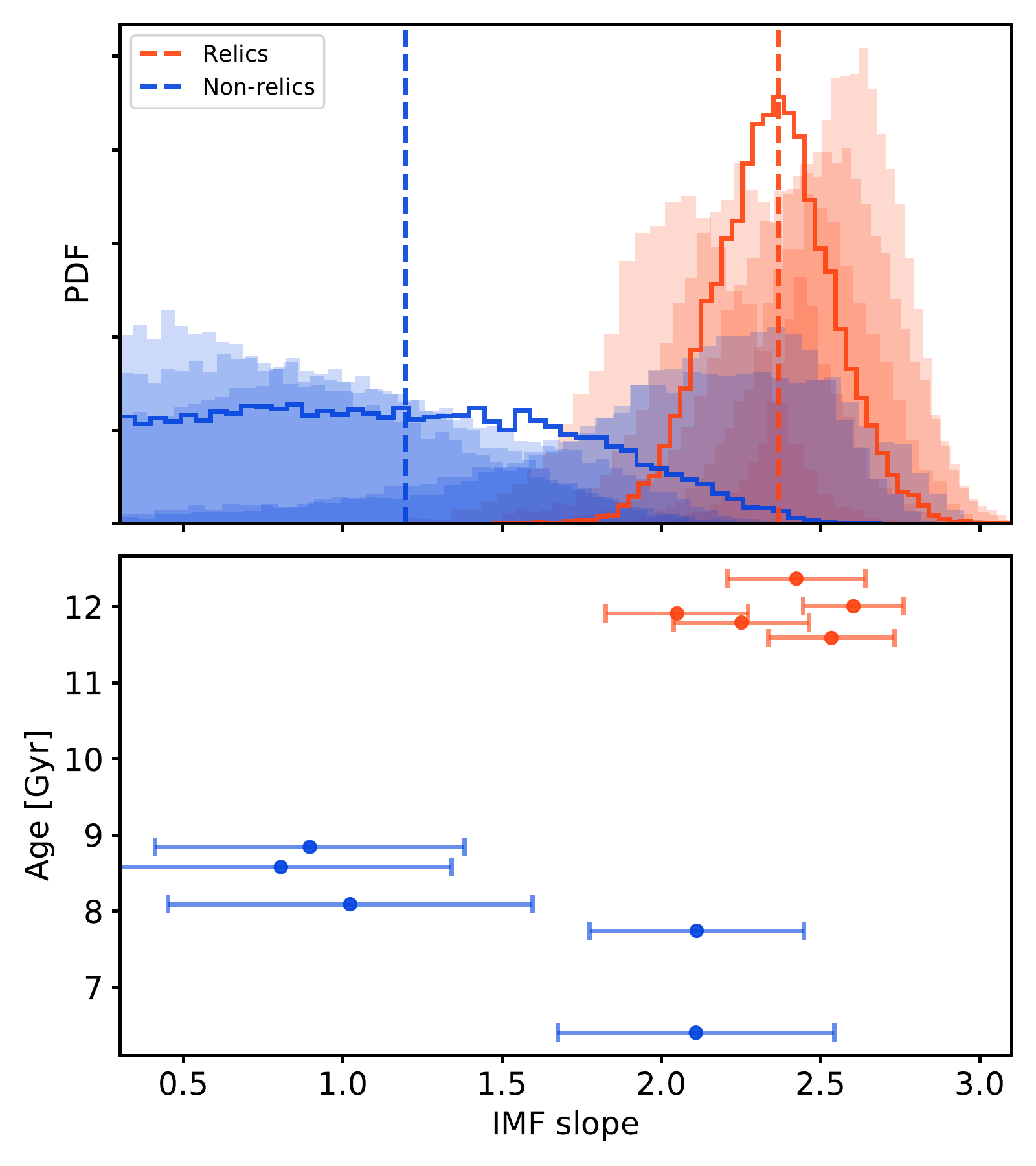}
  \caption{Bootstrapped IMF measurements. Each histogram in the top panel shows the posterior distribution resulting from bootstrapping the stacked spectra of relic (red) and non-relic (blue) galaxies. For each sample, there are five possible combinations of four individual spectra. In the relic sample, all measurements are clustered around bottom-heavy IMF slopes, whereas for the non-relic sample IMF slopes are closer to the Milky Way, value except for the two youngest spectra (bottom panel).}
  \label{fig:imf_boot}
\end{figure}

While Fig.~\ref{fig:imf_pdf} demonstrates the robustness of our measurements against model systematics, the range of possible solutions, in particular for the non-relic sample, is a consequence of the model uncertainties rather than of the actual cosmological evolution that might have differentiated relics from non-relics. In order to better assess the origin of the differences between the two samples, Fig.~\ref{fig:imf_boot} shows the posterior distributions resulting from bootstrapping the samples of relic and non-relic galaxies, excluding a different galaxy from the analysis in each realization. The bottom panel in Fig.~\ref{fig:imf_boot} represents the mean age of these bootstrapped realizations as a function of the measured IMF slope. Bootstrapping across relics and non-relics allows us to probe the diversity of the stellar population properties within both samples.

As revealed by Fig.~\ref{fig:imf_boot}, the posterior distributions of relic galaxies are all clustered around relatively steep IMF slope values, indicating that the underlying stellar populations in this sample are rather homogeneous. On the contrary, two of the bootstrapped stacked spectra of the non-relic sample exhibit IMF values that are marginally consistent with the bottom-heavy IMF slopes of the relic sample, while the rest of the measurements points towards a more Milky Way-like IMF slope. Interestingly, these two realizations with the steepest IMF slope are also those corresponding to the, on average, youngest stellar populations (bottom panel in Fig.~\ref{fig:imf_boot}). From a modelling perspective, our SSP assumption is more reliable for stellar populations with less extended formation histories, as it is the case of those realizations with a Milky Way-like IMF slope (i.e., with older ages). Moreover, with five objects in each stack, our results can be particularly sensitive to outliers and systematics on the individual observed spectra.

\subsection{Mass-to-light ratios and mismatch parameter}
A change in the low-mass end slope of the IMF does not only alter the strength of particular spectral features as shown in Fig.~\ref{fig:fitting}, but it has also an impact on the expected mass-to-light ratio $M/L$. In fact, the agreement between stellar population-based and dynamically-measured IMF variations in nearby massive galaxies remains as one of the strongest arguments in favor of a non-universal IMF beyond the Milky Way \citep[e.g.][]{Treu+10,Spiniello+11,Cappellari+12,Lyubenova+16,Smith2020ARA&A}.

Hence, to facilitate a comparison with dynamical-based studies, it is useful to quantify the expected $M/L$ values for our two samples of relic and non-relic galaxies. Although we note, however, that stellar population-based predictions for the $M/L$ are heavily dependent on the assumed IMF parametrization since the effect of stellar remnants and very-low mass star to the observed spectra is negligible but they can dominate the mass budget. Assuming a broken power-law IMF parametrization \citep[the so-called {\it bimodal} IMF shape in the MILES models notation,][]{Vazdekis96}, our fiducial stellar population model predicts the following $M/L$ values in the SDSS $r$-band

\begin{eqnarray*}
(M/L)_{\mathrm{relics}} = 4.61 \pm 0.59 \\
(M/L)_{\mathrm{non-relics}} = 3.06 \pm 0.45 
\end{eqnarray*}

The effect of a non-universal IMF is often measured in terms of the mismatch parameters $\alpha$, which corresponds to the ratio between the {\it measured} $M/L$, assuming a variable IMF, over the $M/L$ inferred assuming a Milky Way-like IMF. In our samples of relics and non-relics, the values above translate into the following $\alpha$ values for relics and non-relics

\begin{eqnarray*}
  \alpha_{\mathrm{relics}} = 1.59 \pm 0.18 \\
  \alpha_{\mathrm{non-relics}} = 1.10 \pm 0.10 
\end{eqnarray*}

The predicted differences in the mismatch parameter between relics and non-relics are a direct consequence of the measured change in the IMF. We have tested the robustness of these $M/L$ predictions by calculating the $\alpha$ parameter combining all the tests shown in Fig.~\ref{fig:imf_pdf} into a single posterior distribution. In this extreme case, we obtain $\alpha_{\mathrm{relics}} = 1.61 \pm 0.22$ and $\alpha_{\mathrm{non-relics}} = 1.12 \pm 0.13 $, demonstrating that the predicted values for the mismatch parameter of relics and non-relics are indeed robust against modelling systematics, and that they are consistently different.

\section{Discussion and conclusions}

In this paper we have used the relic confirmation from the \citetalias{Spiniello+21} to build two stacked 1D ($\sim60$ per \AA) spectra: one from 5 relics and one from 5 non-relics.  Relic and non-relic galaxies have all similar stellar masses, sizes, integrated velocity dispersions, metallicities and [Mg/Fe] abundances, but objects in the non-relic sample have formed over a longer period of time, i.e, they exhibit more extended star formation histories and therefore younger integrated light-weighted ages.

Using the FIF technique, we have run a detailed stellar population analysis on the UVB+VIS stacked spectra, convolved to the resolution of the MILES SSPs, computing velocity dispersion, stellar population parameters and IMF slope. Our measurements suggest that the IMF in relic galaxies systematically differs from that of the non-relics. In particular, relics host an excess of low-mass stars compared to both non-relic galaxies and the Milky Way standard. On the contrary, the optical spectrum of our sample of non-relic galaxies is consistent with Milky Way-like IMF slope. These differences in the IMF between relic and non-relic galaxies are robust against model systematics and have a direct impact on the expected mass-to-light ratio of both samples. In particular, assuming a broken power-law IMF parametrization, the predicted mismatch parameters  $\alpha$ or relics and non-relics are $1.59 \pm 0.18$ and $1.10 \pm 0.10$, respectively.

When bootstrapping through the individual galaxies included in the stacked data, non-relic galaxies show a wider range of posterior distributions, likely reflecting the more heterogeneous SFHs of these galaxies. Interestingly, the youngest of these bootstrapped spectra of non-relic galaxies are those suggesting steeper IMF slope values, marginally consistent with the rather homogeneous posterior distributions of our sample or relic galaxies (see details above).

This is the first time that the IMF is measured in relics outside the local Universe. Hence, these results add important observational evidence in support of the scenario according to which stars formed during a quick and violent starburst early-on in cosmic time, are distributed with a bottom-heavy IMF \citep{Martin-Navarro+15_IMF_relic,Smith2020ARA&A, Barbosa21_letter}. These stars contribute to almost the totality of the stellar populations in relics, that therefore have a very dwarf-rich IMF, which also stays rather constant with radius \citep{Martin-Navarro+15_IMF_relic,Ferre-Mateu+17}. These old stars dominate the light budget in the innermost regions of massive normal-sized ETGs, where spatial gradients in the IMF have been reported, with a bottom-heavy slope in the centre (e.g., \citealt{Martin-Navarro+15_IMF_variation, Sarzi+18, Parikh+18, LaBarbera+19, Barbosa21_letter}). A general consensus is therefore emerging whereby the non-universality of the IMF slope is due to the formation channel and cosmic-time of the stellar populations. 

Complementary, our measurements of the IMF in the sample of non-relic galaxies point towards a Milky Way-like IMF slope in these objects. This on its own is an interesting result since in the local Universe stellar population properties, and in particular the slope of the IMF, follows tight scaling relations with galaxy stellar velocity dispersion and mass \citep[e.g.][]{Treu+10,Spiniello+12,LaBarbera+13_SPIDERVIII_IMF}. The IMF we measured for our sample of non-relic galaxies is in fact closer to the Milky Way standard than what one would predict from these local scaling relations. As noted above, two of the bootstrapped measurements do suggest a bottom-heavier IMF slope, more consistent with the expectations from the local Universe. However, these two bootstrapped spectra are also the youngest ones, and therefore a steeper IMF slope values may not reflect an actual IMF variation but the unreliability of our SSP modelling assumption to deal with complex and extended star formation histories \citep[e.g.][]{Seidel15}. This issue can be particularly relevant when fitting, as it is our case, temperature-sensitive features like titanium molecular bands.

An immediate question arises if massive non-relic galaxies indeed form their stars following a Milky Way-like IMF: where are the local descendants of these objects? It has been proposed that a fraction of the population of massive galaxies at higher redshifts ends up as the innermost regions of nearby Late-type galaxies \citep[e.g.][]{dlRosa16}, suggesting a range of formation pathways and star formation histories for (old) spheroidal structures in galaxies \citep[e.g.][]{Luca21}. Such a scenario would also be consistent with the very mild IMF variations observed in the bulge of both M31 and the Milky Way \citep{Wegg17,LB21}. The steep IMF value observed in massive relic galaxies would be therefore the result of very extreme star-formation conditions in these objects \citep[e.g.][]{Chabrier_2014}. Surveys at moderate redshifts like \INSPIRE\, will be key to further explore these ideas, both constraining the evolution in the number density of massive galaxies with different star formation histories and by providing precise stellar population measurements across a range of galaxy properties.

In the near future, we plan to extend this study to more objects, taking advantage from the entire \INSPIRE\, catalogue. We will attempt to measure the IMF slope from some of the individual galaxy spectra with high SNR too. Increasing the sample size will also be key to better understand the differences in the IMF between relics and non-relics (see Fig.~\ref{fig:imf_boot}). Finally, we will attempt to extend the spectra fitting to the near infrared (up to $\sim1.5 \ \mu m$) taking advantage from newly developed empirical stellar population models, covering from ultraviolet wavelengths to the infrared regime \citep{Benny16,Vazdekis16,Verro22}. 

\section*{Acknowledgements}
We would like to thank the referee for a very constructive and insightful discussion. IMN and AFM acknowledge support from grant PID2019-107427GB-C32 from the Spanish Ministry of Science and Innovation and from grant ProID2021010080 and CEX2019-000920-S in the framework of Proyectos de I+D por organismos de investigaci\'on y empresas en las \'areas prioritarias de la estrategia de especializaci\'on inteligente de Canarias (RIS-3). FEDER Canarias 2014-2020. CS is supported by an `Hintze Fellowship' at the Oxford Centre for Astrophysical Surveys, funded through generous support from the Hintze Family Charitable  Foundation. 
GD acknowledges support from CONICYT project Basal AFB-170002.

This research made use of Astropy, a community-developed core Python package for Astronomy \citep{Astropy13,Astropy18}, and of the Numpy \citep{Numpy}, SpectRes \citep{Spectres}, and Matplotlib \citep{matplotlib} libraries.

\section*{Data Availability}
UVB and VIS spectra for the  \INSPIRE\, DR1 objects used here are publicly available via the ESO 
Archive Science (https://doi.org/10.18727/archive/36). The NIR spectra will be released as part of the DR2 (D'Ago et al., in prep.)



\bibliographystyle{mnras}





\label{lastpage}
\end{document}